\documentclass[conference]{IEEEtran}
\IEEEoverridecommandlockouts
\usepackage{cite}
\usepackage{amsmath,amssymb,amsfonts}
\usepackage{algorithmic}
\usepackage{graphicx}
\usepackage{textcomp}
\usepackage{xcolor}
\def\BibTeX{{\rm B\kern-.05em{\sc i\kern-.025em b}\kern-.08em
    T\kern-.1667em\lower.7ex\hbox{E}\kern-.125emX}}
\begin{document}

\title{Cultivating Online: Question Routing in a Question and Answering Community for Agriculture}

\author{\IEEEauthorblockN{1\textsuperscript{st} Xiaoxue Shen}
\IEEEauthorblockA{\textit{China Agricultural University} \\
\textit{college of Information and} \\
\textit{Electrical Engineering}\\
Beijing, China \\
xiaoxueshen@cau.edu.cn}
\and
\IEEEauthorblockN{2\textsuperscript{nd} Liyang Gu}
\IEEEauthorblockA{\textit{China Agricultural University} \\
\textit{college of Information and} \\
\textit{Electrical Engineering}\\
Beijing, China \\
guliyang@cau.edu.cn}
\and
\IEEEauthorblockN{3\textsuperscript{rd} Adele Lu Jia}
\IEEEauthorblockA{\textit{China Agricultural University} \\
\textit{college of Information and} \\
\textit{Electrical Engineering}\\
Beijing, China \\
ljia@cau.edu.cn}
}

\maketitle

\footnote{Extension of this work is currently under review and details will be updated soon.}\begin{abstract}
Community-based Question and Answering (CQA) platforms are nowadays enlightening over a billion people with crowdsourced knowledge. A key design issue in CQA platforms
is how to find the potential answerers and to provide the askers
timely and suitable answers, i.e., the so-called question routing
problem. State-of-art approaches often rely on extracting topics
from the question texts. In this work, we analyze the question
routing problem in a CQA system named Farm-Doctor that is
exclusive for agricultural knowledge. The major challenge is that
its questions contain limited textual information.

To this end, we conduct an extensive measurement and obtain
the whole knowledge repository of Farm-Doctor that consists of
over 690 thousand questions and over 3 million answers. To remedy the text deficiency, we model Farm-Doctor as a heterogeneous
information network that incorporates rich side information and
based on network representation learning models we accurately
recommend for each question the users that are highly likely to
answer it. With an average income of fewer than 6 dollars a day,
over 300 thousands farmers in China seek online in Farm-Doctor
for agricultural advices. Our method helps these less eloquent
farmers with their cultivation and hopefully provides a way to
improve their lives.
\end{abstract}


\section{INTRODUCTION}

Community-based Question and Answering (CQA) systems have become popular knowledge sharing platforms where users get answers for the questions they raised.
They have received great attention both in industry and in academia \cite{Chen2016Question,Yang2013CQArank}. % ¼ÓÒýÓÃ
One of the most important goals of CQA systems is to provide an asker with a suitable answer in the shortest possible time, i.e., the so-called \textit{question routing} problem.
In contrast to previous works \cite{Li2010Routing,Li2011Question,Yang2013CQArank}
that focus on general CQA websites such as Quora and Yahoo! Answers, in this work, we analyze the question routing problem in a CQA platform named Farm-Doctor that is exclusive for agricultural knowledge.

In China, over 300 thousand rural resident, although with limited income (on average less than 6 dollars a day), managed to connect to the internet and seek online in Farm-Doctor for agricultural knowledge. Accurate question routing will provide timely advices for their cultivation and potentially improve their lives. However, question routing in Farm-Doctor faces a major challenge, i.e., the limited textual information problem. 

In general CQA platforms, most questions are described in natural languages and question routing is often performed through extracting topics from the rich textual information\cite{Li2010Routing,Yang2013CQArank}. In contrast, users in Farm-Doctor raise their questions mostly through pictures, along with simple questions like \textit{which is the problem?} and \textit{what should I do?} Due to the lack of textual information, topic models that are widely used in CQA platforms are not applicable to the case of Farm-Doctor. On the other hand, although image recognition has received great attention and success both in industry and in academia,
efficient tools on identifying crop diseases and even crops are still missing. As a consequence, it is difficult to infer the topics from the questions (texts or images) alone. To solve this
problem, in this work we incorporate rich side information and model CQA platforms as a heterogeneous information network (HIN), and based on network representation learning (NRL)
models, we conduct, to the best of our knowledge, the first analysis of the question routing problem without using textual information.

Our analysis of Farm-Doctor mainly consists of three parts:

\textbf{Measuring Farm-Doctor.} In this work, we have obtained the whole knowledge base of Farm-Doctor. Our dataset covers all the 697,695 questions raised before April 21st, 2018, along
with the information on the associated 3,179,333 answers, 438 crops and 305,359 users. The information we obtained include not only the basic question characteristics but also user activities such as who raised/answered which question at what time, which crop is tagged in which question and
is interested to which user. The dataset is publicly available through requests to the first author.

\textbf{Characterizing question and answer dynamics.} We first provide an analysis on the scale and the characteristics of the question repository in Farm-Doctor. We examine the number
of answers received by each question and the timeliness of the answers. We find that questions in Farm-Doctor normally attract a few answers shortly after they are raised, but as
time passes by they no longer receive any attention. Then, we analyze the user activities in terms of the number questions they raised and answered. We find a highly skewed activity
level of the users, with a small number of users raising and answering a large number of questions. These results all indicate the need of a proper question routing method in Farm-Doctor, so that questions will get more answers and hopefully that less active users will be encouraged to answer
the questions personalized recommended to them.

\textbf{Question routing.} To tackle the limited textual information problem in Farm-Doctor, we build a HIN model based on a variety of relationships and we adopt heterogeneous NRL models to learn the low-dimensional embeddings of the questions, the users, and the crops. Taking the learned
representations as input features, we build machine-learned classifiers to recommend timely and accurately, for the newly posted questions, the potential users that most likely will
answer the questions.

The main contributions of this work are as follows:

\begin{itemize}
  \item We obtain the whole knowledge repository of FarmDoctor (until April 2018) that contains information of 305,359 users, 697,695 questions, and 3,179,333 answers (Section 2). Our dataset is publicly available upon request to the first author.
  \item We analyze the basic characteristics of the question repository and the user activities in Farm-Doctor (Section3).
  \item We propose a heterogeneous graph model that incorporates a variety of relationships among users, questions and crops. Based on this model, we adopt NRL methods to, for each question, accurately predict the potential answerers (Section 4).
\end{itemize}

\bibliographystyle{IEEEtran}
\bibliography{sample-bibliography}

\end{document}